\documentclass[10pt,journal,cspaper,compsoc]{IEEEtran}

\usepackage[labelfont=bf]{caption}
\usepackage{graphicx}
\usepackage{amssymb}
\usepackage{amsmath}
\usepackage{colortbl}
\usepackage{setspace}
\usepackage{color}
\usepackage{url}
\usepackage[lofdepth,lotdepth]{subfig}
\usepackage{multirow}
\usepackage{threeparttable}
\usepackage{tikz}
\usepackage{bbding}
\usepackage{etex}
\usepackage[all]{xy}
\usepackage{indentfirst}
\usepackage{balance}

\newcommand{\descr}[1]{\vspace{0.25cm} \noindent \textbf{#1}}
\newcommand{\alg}[1]{\vspace{0.25cm} \noindent \textbf{\em #1 }}

\def\naive{na\"{i}ve}

\newcommand{\F}{\mathbb{F}}
\newcommand{\M}{\mathcal{M}}

\newcommand{\D}{\ensuremath{\mathcal{D}}}
\newcommand{\Q}{\ensuremath{\mathcal{Q}}}
\newcommand{\N}{\ensuremath{\mathcal{N}}}
\newcommand{\Z}{\mathbb{Z}}
\newcommand{\G}{\mathbb{G}}
\newcommand{\adv}{\ensuremath{{\mathcal A}}}

\newcommand{\remove}[1]{}
\newcommand{\spa}{\mbox{ }}
\newcommand{\modified}{\textcolor{black}}

\begin{document}
\title{Extended Capabilities for a Privacy-Enhanced Participatory Sensing Infrastructure (PEPSI)}

\author{Emiliano De Cristofaro$^{1}$ and Claudio Soriente$^2$ \vspace{0.2cm}\\
{\normalsize $^1$ PARC\\ $^2$  ETH Z\"{u}rich, Switzerland}}

\date{}

\maketitle

\begin{abstract}
Participatory sensing is emerging as an innovative computing paradigm that targets
the ubiquity of always-connected mobile phones and their sensing capabilities.
In this context, a multitude of pioneering applications increasingly carry out pervasive collection and dissemination of information and environmental data, such as, traffic conditions, pollution, temperature, etc.
Participants collect and report measurements from their mobile devices and entrust them to the cloud to be made available to applications and users. Naturally, due to the personal information associated to the reports (e.g., location, movements, etc.), a number of privacy concerns need to be taken into account prior to a large-scale deployment of these applications.
Motivated by the need for privacy protection in Participatory Sensing, this work presents PEPSI: a Privacy-Enhanced Participatory Sensing Infrastructure.
We explore realistic architectural assumptions and a minimal set of formal requirements aiming at protecting privacy of both data producers and consumers.
We propose two instantiations that attain privacy guarantees with
provable security at very low additional computational cost and almost no extra communication overhead.
\end{abstract}

\section{Introduction}\label{sec:introduction}
In the last decade, a number of cloud-based services have emerged as the key component for
information sharing in online communities.
More recently, technology has taken a turn toward context-aware, ubiquitously-available
information contributed by individuals from their mobile devices.
In particular, \textbf{\em participatory sensing} \cite{participatory}
entails the widespread availability of always-on, always-connected mobile devices
as well as their  sensing capabilities.
Participatory sensing allows cloud-based services to harvest and share dynamic information about environmental trends,
such as ambient air quality~\cite{atmosphere},
parking availabilities~\cite{trappe},
earthquakes~\cite{quake},
consumer pricing information in offline market~\cite{fuel,mobishop}, and so on.
In the last few years, participatory sensing initiatives have multiplied, ranging from research
prototypes~\cite{trappe,sensor-planet} to deployed systems~\cite{atmosphere}.

A typical participatory sensing infrastructure involves the following parties:
\begin{itemize}
\item \textbf{Sensors}, installed on smartphones
or other wireless-enabled devices, emit data reports and form the basis of any participatory sensing infrastructure.
\item \textbf{Sensor Carriers}  are usually users who carry their smartphones, but
they could also be vehicles, animals or any other mobile entity equipped with a portable sensing device that has network connectivity.
Hereafter, we denote a sensor and its carrier as \textbf{\em Mobile Node}.
\item \textbf{Network Operators} manage the network used to collect and deliver reports, e.g., maintaining a WiFi, GSM, or 3G network infrastructure.
\item \textbf{Queriers} subscribe to specific information collected in a participatory sensing application (e.g.,
``{\em temperature readings from all sensors in San Francisco, CA}'') and obtain corresponding data reports.
\item \textbf{Service Providers} are cloud-based services that allow effective sharing of information between mobile nodes and queriers.
Since mobile nodes and queriers have no mutual knowledge, service providers are key to participatory sensing applications.
Data is reported and searched by its ``nature'' (e.g., pollution level in Central Park, New York) and service providers
are responsible for data collection and dissemination to interested queriers.
\end{itemize}

The research community is currently proposing platforms for application developers~\cite{prism} and devising innovative business models,
based on incentive mechanisms, for the capitalization on sensed data~\cite{estrin2}.
Unfortunately, security and privacy issues in this emerging computing paradigm have not always been effectively addressed.
One possible reason is that participatory sensing seemingly shares several features with Wireless Sensor Networks (WSNs),
a relatively more mature research field where a number of effective means to guarantee security have been proposed \cite{chen09cst}.
However, there is a number of crucial differences between these two areas.
First, as opposed to low-cost, resource-constrained motes in WSNs, \emph{sensors} in
participatory sensing applications are high-end mobile devices (e.g., smartphones).
They have considerable computational power and their batteries can be easily recharged.
Also, privacy concerns in WSNs are usually not so severe as a common assumption
is that the network operator owns and queries all sensors.
This is not the case in most participatory sensing scenarios.
For instance, consider the case of an acoustic pollution monitoring application: in a WSN scenario, the main stakeholder, e.g., the Environmental Protection Agency,
would set up a network to sense noise and would regularly query its sensors to collect measurements.
The same application in a participatory sensing infrastructure, would employ noise sensors
embedded in users' smartphones to collect noise readings and report them to a cloud provider, which would
forward reports to interested queriers. Clearly, each report entails a significant amount of private information that reveals,
e.g., user's position and activity, thus, raising severe privacy concerns.

\subsection{Motivation \& Contributions}
In several participatory sensing settings, there is a tension between privacy protection and user participation.
On the one hand, the success of any application relies on context-aware information sensed (and voluntarily shared)
by users from their mobile devices. On the other hand, this sharing often discloses personal information and presents a fundamental obstacle
to large-scale deployment of applications---arguably, users may decide to opt-out if they feel that their privacy is endangered.
The number and the heterogeneity of entities involved in participatory sensing prompts a range of new
formidable privacy challenges that must be carefully addressed.
Prior work has proposed a few solutions to protect privacy of user locations and reports.
However, it has often introduced unrealistic assumptions and failed to provide provable security.
On the contrary, we aim at a {\em formal treatment} of participatory sensing.
We investigate realistic architectural assumptions and a minimal set of formal requirements
intended to protect privacy of both data producers (i.e., mobile nodes)  and data consumers (i.e., queriers).
We present two instantiations that attain privacy guarantees with provable security,
at very low additional computational cost and almost no extra communication overhead.

\subsection{Paper Organization}
The rest of this paper is structured as follows.
Section~\ref{sec:related} surveys related work and highlights its limitations. Then,
Section \ref{sec:blocks} presents main building blocks, parties and operations and the privacy desiderata for a
privacy-enhanced participatory sensing application.
Section \ref{sec:pepsi} introduced an instantiation of PEPSI based on Identity-Based Cryptography, while Section \ref{sec:pri-an} analyzes
its privacy guarantees.
Section \ref{sec:oprfpepsi} presents another instantiation of PEPSI based on Oblivious Pseudo-Random Functions.
Performance of both instantiations are discussed in Section \ref{sec:perf}.
Finally, the paper concludes in Section \ref{sec:conclusion}.

\section{Related Work}\label{sec:related}
Participatory sensing has attracted, in recent years, great interest from the research community.
Security and privacy challenges have been widely discussed in
\cite{shilton},~\cite{kapadia2009},~\cite{manulis},~\cite{christin11jss} but none of them
has proposed actual solutions.
To the best of our knowledge, AnonySense~\cite{mobisys08}
(later extended in \cite{shin11tmc} and~\cite{comcom})
is the only result to address privacy-related problems, hence, it is most related to our work.
AnonySense leverages Mix Network techniques~\cite{MixNet} and provides $k$-anonymity~\cite{k-anonymity},
while \cite{comcom} shows how to modify the original AnonySense,
to achieve $l$-diversity \cite{l-diversity}.

Both \cite{mobisys08} and \cite{comcom} guarantee
report integrity using group signatures
(i.e., all sensors share the same group key to sign reports) and
provide limited  confidentiality, as reports are encrypted under the public key of
a \emph{Report Service},
a trusted party responsible for collecting reports and distributing them to queriers.

There is also additional research work that focuses on slightly related problems.
For instance, \cite{caceres09mobiheld} argues that privacy issues can be addressed if
each user has access to a private server
(e.g., a virtual machine hosted by a cloud service)
and uses it as a proxy between her {\em sensors} and applications requesting her data.
Nevertheless, the feasibility of the approach in large scale participatory sensing applications  would be severely limited by cost and availability of per-user proxies.

Authors of \cite{ganti2008} show how to protect user anonymity while computing
community statistics on time-series data. Their approach leverages data perturbation
and can only be used in closed community scenarios, where an empirical data distribution
is known a priori.
Data perturbation is also at the basis of \emph{differential privacy} \cite{dwork06icalp}
that aims to provide the maximal accuracy of responses to queries issued to a statistical database,
while preserving the privacy of single records.
Similarly, \cite{infocom10} targets privacy-preserving data aggregation,
e.g., computation of sum, average, or variance.
Our research does not focus, at the moment, on aggregates or statistics
of the reported data, therefore, aggregation or perturbation techniques are not applicable to our setting.

Other proposals, such as~\cite{dua2009}
and~\cite{gilbert2010}, aim at guaranteeing integrity and authenticity of user-generated contents,
by employing Trusted Platform Modules (TPMs).

\subsection{Limitations of prior work}

The goal of our work is to afford \emph{provable privacy guarantees} in participatory sensing applications.
To this end, we now discuss in details limitations and open problems of prior work in the area.

\descr{Limited communication infrastructure.}
Most relevant work on privacy for participatory sensing (e.g., \cite{mobisys08,comcom,infocom10}) assumes
that measurements are reported via WiFi networks.
In particular, \cite{mobisys08,comcom} use standard MAC-IP address recycling techniques to guarantee user unlinkability between reports with respect to WiFi access points.
Despite extensive research on privacy, anonymity, and unlinkability in WiFi networks \cite{levente}, assuming WiFi as the underlying network, poses
severe limitations on the scope of participatory sensing applications.
This is because an ubiquitous presence of open WiFi networks is not realistic today nor anticipated in the next future, while participatory sensing requires always-on, always-connected devices.
Actually, the majority of existing participatory sensing applications operate from smartphones
and use cellular networks to upload reports~\cite{bike,atmosphere,trappe}.
Unfortunately, current technology does not allow to apply WiFi anonymization techniques to cellular networks. In fact, in cellular networks devices are identified through their
International Mobile Subscriber Identity (IMSI), and ID recycling---besides being impossible with  current technologies---would lead to denial of service (e.g., the device would not receive incoming calls for its original ID).
Moreover, the regular usage of cellular networks (e.g., including incoming/outgoing phone calls), as well as heartbeat messages exchanged with the network infrastructure, irremediably reveal device's location to the network operator.

\descr{No provable privacy.}
User privacy in previous work (e.g., \cite{mobisys08,comcom})
relies on Mix Networks~\cite{MixNet},
an anonymizing technique used to de-link
submitted reports from their origin, before they reach applications.
In other words, a Mix Network acts as an anonymizing proxy and forwards
user reports only when the set of received reports satisfies a system-defined criteria.
Privacy metrics such as $k$-anonymity \cite{k-anonymity} or $l$-diversity \cite{l-diversity} have been defined to characterize privacy through Mix Networks.
For example, a Mix Network that provides $k$-anonymity ``batches'' reports so that
it is not possible to link a given report to its sender, among a set of $k$ reports.
Clearly, anonymity is not guaranteed but it rather depends on the  number of reports received and ``mixed'' by the Mix Network.
Moreover, there could be scenarios where a relatively long time could pass before the desired level of anonymity is reached (when ``enough'' reports have been collected).

\descr{Multiple Semi-Trusted Parties.}
Trust relations are difficult to define and set up in scenarios with multiple parties.
Hence, it is advisable to minimize the number of trusted parties and the degree to which they are trusted.
Available techniques to protect privacy in participatory sensing
often involve many semi-trusted independent parties, that are always assumed not to collude.
Anonysense \cite{mobisys08}, besides Mobile Nodes,
Registration Authority, and WiFi Access Points, also assumes the presence and the non-collusion of
a Task Service (used to distribute tasks to users), a Report Service (to receive reports from sensors),
and several Mix Network nodes (i.e., a trusted anonymizing infrastructure).
The assumption of multiple non-colluding parties raises severe concerns regarding
its practicality and feasibility.
It appears difficult to deploy all of the parties in a real world setting where entities
provide services only in exchange of some benefit.
For instance, it is not clear how to deploy the Task and the Report services as two separate entities having no incentive to
collude.

\section{Preliminaries}\label{sec:blocks}
Aiming at a cryptographic treatment of privacy in participatory sensing,
we now formalize entities and protocols
involved in a privacy-enhanced participatory sensing infrastructure.
Before, we introduce some cryptographic background.

\subsection{Cryptographic Background}
We now provide an overview of building blocks used in our work,
namely, Identity Based Encryption (IBE) and Oblivious Pseudo-Random Functions (OPRF).

\subsubsection{Identity Based Encryption}\label{sec:sub:ibe}

Identity Based (IB) cryptography represents an effective alternative to traditional public key cryptography,
as it does not rely on Public Key Infrastructure responsible to bind public key to parties.
In an IB cryptographic scheme, the identities of the parties ``are'' their public keys.
In other words,
any string can serve as a public key; corresponding private keys are
managed and issued by a trusted authority, referred to as Private Key
Generator (PKG). The latter is only active when issuing new keys.

We take advantage of IB cryptography to afford effective
communication between mobile nodes and queriers.
In particular, we use the IB Encryption scheme of Meiklejohn et al. \cite{meiklejohn11usenix} that was inspired by the IB scheme by Boneh and Franklin~\cite{IBE}.
The former is composed by four algorithms:
\texttt{Setup} is run by a trusted authority to
set up system parameters and a master secret key;
\texttt{BlindExtract} is an interactive protocol between an user and the authority wherein the user obtains a secret key for an identity of her choice, while the authority learns nothing about the identity
picked by the user;
\texttt{IBEnc} encrypts a message for a recipient specified by an arbitrary identity;
\texttt{IBDec} allows for decryption of ciphertext produced for a given identity, given the corresponding secret key.
\begin{itemize}
\item \texttt{Setup}$(1^\lambda)$.
Given a security parameter $\lambda$, generate two groups group $\G,\G_T$ of prime order $q$ with generator $g$ and an associated symmetric bilinear map $e: \G\times \G \rightarrow \G_T$.
Message space is set to $\M=\{0,1\}^n$ for some value $n$.
Pick random $x_1, x_2 \in {\F_q^*}$ and set $X_1=g^{x_1}, X_2=g^{x_2}$.
Choose two cryptographic hash functions
$H: \{0,1\}^* \rightarrow \G$ and
$H':\{0,1\}^* \rightarrow \M$.
Secret master key is set to $msk=(x_1,x_2)$, while public key is set to $pk=(q,\G,\G_T,g,e,\M,X_1,X_2,H,H')$.
\item \texttt{BlindExtract}(\texttt{User}$(pk,id)\leftrightarrow$\texttt{Auth$(msk)$}).
Given a string $id \in \{0,1\}^*$, the user picks a random $r\leftarrow\F_q^*$ and sends
$req=H(id)\cdot g^r$ to the authority. The latter replies with $sk'_1=req^{x_1},sk'_2=req^{x_2}$.
Finally, the user computes $sk_1=\frac{sk'_1}{X_1^r}, sk_2=\frac{sk'_2}{X_2^r}$ and sets
the secret key for $id$ to $sk_{id}=(sk_1,sk_2)$.
Keys are properly formed if $e(sk_1,g)=e(H(id),X_1)$ and $e(sk_2,g)=e(H(id),X_2)$.
\item \texttt{IBEnc}$(pk,id,m)$.
Pick random $r\leftarrow\F_q^*$ and compute $Z_1=e(H(id),X_1)^r, Z_2=e(H(id),X_2)^r$.
Then, compute $h=H'(id,g^r,Z_1,Z_2)$ and output $c=(g^r,h\oplus m)$.

\item \texttt{IBDec}$(pk,sk_{id},c)$.
Parse $c=(c_1,c_2)$ and $sk_{id}=(sk_1,sk_2)$.
Compute $Z_1=e(sk_1,c_1), Z_2=e(sk_2,c_1)$ and $h=(id,c_1,Z_1,Z_2)$.
Output $m=h\oplus c$.
\end{itemize}

The above scheme is IND-ID-CPA secure under the Bilinear Diffie-Hellman (BDH) assumption.
Further it enjoys selective-failure blindness \cite{blind-ibe} and one-more indistinguishability \cite{meiklejohn11usenix}.

\subsubsection{Oblivious Pseudo-Random Functions}\label{sec:sub:oprf}
A pseudo-random function (PRF) family $\Phi$ is a collection of efficiently
computable functions such that, fixed one function in $\Phi$ at random,
and given the function outputs for a number of arbitrary inputs,
no efficient algorithm has a non-negligible advantage
in telling if the outputs belong to one function randomly chosen from $\Phi$ or
a true random function.
A PRF $f$ is a function that takes two inputs: a variable $x$ and a function index
$s$, and outputs $f_s(x)$.

We then look at two-party computation of PRFs, and in particular to
{\em Oblivious PRFs} (OPRFs). An OPRF is a two-party protocol, between a sender on private input $s$,
and a receiver on private input $x$.
At the end of the interaction, the receiver obtains $f_s(x)$
while the sender learns nothing.
OPRFs have been introduced by Freedman, et al.~\cite{freedman05tcc}, based on Naor-Reingold PRF~\cite{naor97focs}.

\begin{figure}[t!]
\centering
\includegraphics[angle=0, width=90mm, height=60mm]{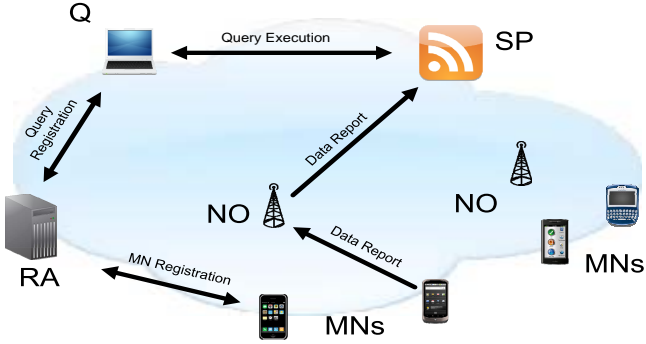}
\caption{Privacy-Enhanced Participatory Sensing Infrastructure: Mobile Nodes (MNs) register
to the Registration Authority (RA) and, subsequently, report sensed data to a Service Provider (SP), through the Network Operator (NO).
Queriers (Q), after registering to RA, subscribe to queries offered by the SP and receive corresponding
reports.}
\label{fig:architecture}
\end{figure}

\subsection{Infrastructure}\label{sec:infrastructure}
We envision a participatory sensing infrastructure, that we call PEPSI (Privacy-Enhanced
Participatory Sensing Infrastructure) composed by the following entities:

\descr{Mobile Nodes (MNs).} They are portable computing devices with sensing capabilities (i.e.,
equipped with one or more {\em sensors}) and with access to a cellular network.
They are carried by people or attached to mobile entities. %
We assume that MNs run on smartphones and that users voluntarily engage into participatory sensing.
We denote with $\mathcal{N}$ a generic mobile node
of a participatory sensing application.

\descr{Queriers.} Queriers are end-users interested in receiving sensor reports in a given
participatory sensing application.
A generic querier is denoted with \Q. %

\descr{Network Operator (NO).} The Network Operator  is responsible for the communication infrastructure.
We assume that the NO maintains, and provides access to, a cellular network infrastructure (e.g., GSM or 3G).

\descr{Service Providers (SP).} As mobile nodes and queriers might have no mutual knowledge,
the Service Provider acts as an intermediary between them. That is, SP collects
mobile nodes reading and forwards them to interested queriers.
We envision one or more cloud-based SPs running participatory sensing applications that offer different query types. (For example,
a national service provider, or a commercial entity, may run a pollution monitoring application and define queries to retrieve reports of pollution levels
in different cities). Service provider's duties may include listing available sensing services,
micro-payment, data collection, and notification to queriers.

\descr{Registration Authority (RA).} The Registration Authority  handles the application setup,
as well as the registration of participating parties. In our solutions, the RA
also contributes to privacy protection, by generating cryptographic public parameters, handling the registration of MNs,
and managing queriers' subscription. Note that the Registration Authority is the only additional party required by PEPSI
in comparison with a ``regular'' (i.e., non-private) participatory sensing application.

\subsection{Operations}\label{sec:operations}
We now describe the common operations performed within a privacy-preserving participatory sensing application and the interactions between the entities introduced above.

\descr{Setup.} In this phase, the RA generates all public parameters and its own secret key.

\descr{MN Registration.} 
Users register their sensor-equipped device to the RA and install participatory sensing software.
\modified{
At this time, the Mobile Node may also fetch the list of data reports types for which it will later provide reports.
We assume a public list of report types may be available from either the SP or the RA.
}

\descr{Query Registration.} Queriers approach the appropriate RA and
request an {\em authorization} to query the participatory sensing application, in order to obtain a specific type of
data reports, e.g., {\em ``Pollution level in Manhattan, NY''}.
\modified{
Similar to MN registration, available report types are public, and the list is hosted, e.g., at the RA or at a SP.
}
Next, they may subscribe to one or more (authorized) queries,
by submitting a request to a SP and awaiting for the responses containing the desired readings.
Ideally, only queriers authorized by the RA should receive the desired reports.
Also, no information about query interests should be revealed to the SP nor to the RA.

\descr{Data Report.} Mobile Nodes report to the Service Provider their readings, using the network access
provided by the Network Operator. Ideally, this operation should not reveal to the SP, the NO, or unauthorized queriers
any information about reported data,
such as type of reading (e.g., pollution) or quantitative information (e.g., 35$mg/m^3$ carbon oxide).
Also, the SP and any querier should not learn the identity of the source MN.

\descr{Query Execution.} With this operation, the SP matches
incoming data reports with query subscriptions. Ideally, this should be
done {\em blindly}, i.e., the SP should learn nothing beyond the occurrence
of an (unspecified) match, if any.

\noindent Figure~\ref{fig:architecture} shows our participatory sensing infrastructure.
In the depicted scenario, one may envision that a phone manufacturer (e.g., Nokia) acts as the RA and embeds a given type of sensor (e.g., air pollution meter)
in one or more of its phone models, operated by smartphone users, i.e., MNs.
A service provider (such as Google, Microsoft, Yahoo, or a non-profit/academic organization) %
offers participatory sensing applications (used, for instance,
to report and access pollution data), and acts as an intermediary between queriers and mobile nodes.
Finally, queriers are users or organizations (e.g., bikers) interested in obtaining readings (e.g., pollution levels).

Note that---similar to related work---we do not address
the problem of encouraging mobile phone users to run participatory sensing applications, nor
we focus on business incentives for phone manufacturers or for service providers.
Nonetheless, it is reasonable to envision that queriers are willing to pay small fees (or receive advertisement)
in return to obtaining measurements of interest.

\subsection{Privacy Requirements}
\label{sec:requirements}
Before entering the details of our privacy requirements, observe that
the main purpose of a participatory sensing application is to allow queriers to obtain MNs reports.
While our main goal is to protect the privacy of both data producers and consumers,
any entity registered as querier should be able to receive desired measurements, thus,
techniques to identify legitimate parties before registering them are beyond the scope of our work.

We now define the privacy requirements of PEPSI.
Our definitions here are concise, whereas, formal adversarial games can be found
in Section \ref{sec:pri-an}.
In what follow, we assume honest-but-curious parties, i.e., each party acts
according to the prescribed protocols, however, they might try to infer as much information as possible
from obtained results.

\descr{Soundness.} We say that PEPSI is {\em sound} if, upon subscribing to a query,
a querier in possession of the appropriate authorization obtains the desired readings (if any).

\descr{Node Privacy.}
In a privacy-enhanced participatory sensing application,
node reports should be only available to authorized queriers.
In particular, even if measurements are routed through the service provider, the latter should not learn any information
about reported data.
We say that PEPSI is {\em node-private} if
neither the NO, the SP, nor any unauthorized querier, learn any information
about the type of reading or the data reported by the MN. Also, other MNs
should not learn any information about a given node's reports.

In other words, \emph{only queriers in possession of the corresponding authorization obtain MN's readings}.

\descr{Query Privacy.}
Similar to mobile node reports, query interests provide a considerable amount of personal information.
Our goal is to keep query interests private w.r.t. \emph{all} other parities in the system.
This includes the service provider that should deliver reports to queriers without learning
their interests. Further, while we trust the RA to manage mobile nodes and querier registrations,
we would like it to learn as little information as possible about interests of queriers.

We say that a PEPSI is {\em query-private} if
neither the RA, the NO, the SP,  nor any mobile node or any other querier, learn any information
about the query subscribed by a querier.

\descr{Anonymity, Report Unlinkability, and Location Privacy.}
While a number of proposals attempt to provide anonymity, report unlinkability, and location privacy in mobile applications \cite{krumm09puc}, we
argue that it is not possible to guarantee these features with respect to the Network Operator (NO), thus, we do not consider them in this manuscript.
The NO, in cellular networks, is required to know identity and location of a mobile node at any time (e.g., to provide network connectivity).
Furthermore, current proposals (e.g., \cite{beresford03pervasive}) to protect location privacy in Location-Based Services (LBS) cannot be directly applied to participatory sensing application, as they focus on public databases (e.g., fast-foods or ATMs locations) that users query providing ``a location'' to refine their search. Whereas, in participatory sensing applications, both queriers and mobile nodes include location information in their queries or data reports.

\section{Pepsi Instantiation}\label{sec:pepsi}
We now present our PEPSI instantiation, in accordance to the architectural design of Figure~\ref{fig:architecture} and
that complies with privacy requirements of Section~\ref{sec:requirements}.

\subsection{High Level Description}\label{sec:highlevel}
\modified{In PEPSI, data reports are always labeled using keywords that identify
the nature of the information announced by MNs. Similarly, queriers subscribe
to given queries by specifying the corresponding keywords.
In the rest of the paper, we use the term {\em identifier}, and the notation
$ID$ (or $ID^*$) to identify the data report/query type. Examples of such identifiers
include: ``{\em Temperature in San Francisco, CA}'' or ``{\em Pollution in Manhattan, NY}''.
The list of identifiers -- depending on the application -- can be obtained either from the SP
or the RA. In particular, the RA defines which services (i.e., queries) will be available
for MNs to contribute and for users to query. However, as these identifiers can be public,
they can be downloaded from the SP or any bulletin board.
For ease of presentation, in the rest of the paper, we assume that query identifiers are available at the RA.}

One of the main goals of PEPSI is to hide reports and queries to unintended parties.
Thus, those cannot be transmitted {\em in-the-clear}, but need to be encrypted.
In the rest of this section, we discuss how to achieve, at the same time, (1) secure encryption of reports and queries,
and (2) efficient and oblivious matching performed at the Service Provider.

\descr{Report/Query Encryption.} %
One \naive\ possibility is to let each querier and each mobile node share a unique pairwise secret key and employ a symmetric-key cipher, such as AES~\cite{AES}.
This approach requires queriers and MNs to interact and establish a shared secret.
On the contrary, participatory sensing requires no contact (nor mutual knowledge) between them: that is,
MNs provide reports obliviously of (any) potential receiver. Similarly, queries subscribe to measurements
without knowing the identity of MNs producing reports potentially matching their interests.
Even if we allow interaction between each mobile node and queriers,
we would still need MNs to encrypt reports under each key shared with queriers (recall that MNs do not know
which queriers are interested in their reports).
This would generate a number of ciphertexts  quadratic in the number of measurements.
Alternatively, we could use a public key encryption scheme and provide MNs with public keys of queriers.
Still, scalability would be an issue as each report would be encrypted under the public key of each querier.

\descr{Using Identity-based Encryption.}
PEPSI's main building block is Identity-Based Encryption (IBE), specifically, the construction given by Meiklejohn et al. \cite{meiklejohn11usenix}.
The main advantage in using IBE, as opposed to standard public-key cryptography,
is to enable non-interactivity in our protocol design. This is crucial in participatory
sensing scenarios, where MNs and queriers have no direct communication nor mutual knowledge.
IBE enables asymmetric encryption using any string (``identity'') as a public key.
Recall that, in IBE, anyone can derive public keys from some unique information about recipient's identity.
Private decryption keys are generated by a third-party, called the Private Key Generator (PKG).

Our idea is to use labels (i.e., keywords) that define the type of reports as identities
in an IBE scheme. For examples labels ``Temperature'' and ``San Francisco'' can be used to
derive a unique public encryption key, associated to a secret decryption key.
Mobile Nodes encrypt sensed data using report's labels as the (public) encryption key.
Query registration then consists in obtaining the private decryption keys corresponding
to the labels of interest.
Decryption keys are obtained, upon query registration, from the Registration Authority -- which acts
like a PKG.

\descr{Efficient matching.}
Given encryption/decryption of reports, we still need a way for the Service Provider to match them against queries.
To address this problem, we leverage an efficient tagging mechanisms: Mobile Nodes tag each report with a
cryptographic token that identifies the nature of the report only to authorized Queriers, but does not leak any
information about the report itself. Tags are computed using the same keywords used to derive encryption keys.
Similarly, Queriers compute tags for keywords of interest, using the corresponding decryption keys, and provide them to the Service Provider at query subscription.
Bilinear maps allow us to derive the same cryptographic tag for both a report or a subscription, as long as they are based on the same keywords.
Given our tagging mechanism, the Service Provider is only required to match tags that accompany reports to the ones used by queriers during their subscription.
Any time a match is found, the report is marked for delivery to the corresponding querier.

\descr{PEPSI Overview.}
PEPSI works as follows. The RA runs the
\emph{Setup} algorithm to generate public parameters and secret keys.
In order to pose a query, e.g., identified by $ID^*$, queriers first need to register to the RA
and obtain the corresponding authorization (\emph{Query Authorization}). Then, they
subscribe their queries to the SP (\emph{Query Subscription}). In PEPSI, both processes reveal nothing about queriers' interests.
At the same time, before starting reporting data,
MNs need to authenticate to the RA, and obtain: (i) the identifier $ID$ corresponding
to the type of their reports, and (ii) a token that allows them to announce data (\emph{MN Registration}).
The {\em on-line} part of PEPSI includes two operations:
\emph{Data Report} and \emph{Query Execution}. With the former,
MNs upload encrypted reports to the SP. In the latter, the SP {\em blindly} matches received reports
with queries and forwards (matching) reports to interested queriers. Only
authorized queriers obtain query responses, can decrypt data reports, and retrieve original measurements.
Finally, we let the RA periodically run a \emph{Nonce Renewal} procedure to evict
malicious MNs from the participatory sensing application.\footnote{Techniques
to identify malicious MNs are beyond the scope of this work.}
This procedure is run periodically (e.g., once a week or once a month)
and the new nonce is securely delivered to honest MNs, e.g., using broadcast encryption~\cite{broadcast}.

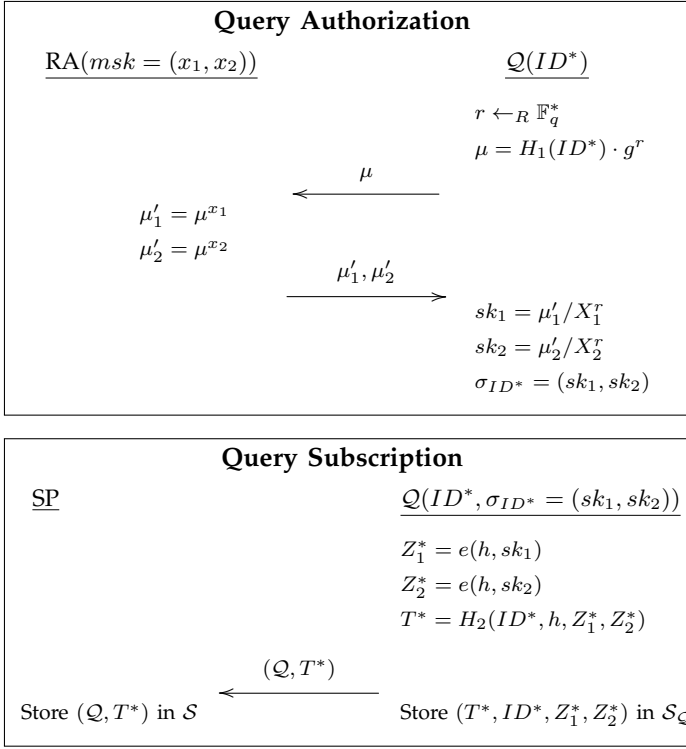
\begin{figure}[t!]
\fbox{\footnotesize
\begin{minipage}[t]{1\columnwidth}
\centering {\normalsize \textbf{Query Authorization}} \vspace{0.2cm}\\
\begin{tabular}{lcl}
\hspace{-0.05cm} \underline{\small RA$(msk=(x_1,x_2))$}\vspace{0.1cm} && \hspace{0.2cm}\underline{\small $\Q(ID^*)$}\vspace{0.2cm}\\
&&\hspace{-0.2cm}$r \leftarrow_R \F^*_q$\vspace{0.15cm}\\
&&\hspace{-0.2cm}$\mu = H_1(ID^*)\cdot g^r$\vspace{-0.1cm}\\
&\hspace{-0.3cm} $\xymatrix@1@=55pt{& \ar[l]^*+<4pt>{}_*+<4pt>{\mu}}$ &\vspace{-0.05cm}\\
\hspace{1.2cm} $\mu'_1=\mu^{x_1}$ \hspace{-0.4cm}\spa &&\vspace{0.15cm}\\
\hspace{1.2cm} $\mu'_2=\mu^{x_2}$ \hspace{-0.4cm}\spa && \vspace{-0.2cm}\\
&\hspace{-0.2cm} $\xymatrix@1@=60pt{\ar[r]^*+<4pt>{\mu'_1,\mu'_2}&}$ \vspace{-0.1cm}\\
&&\hspace{-0.2cm}$sk_1=\mu'_1 / X_1^r$\vspace{0.15cm}\\
&&\hspace{-0.2cm}$sk_2=\mu'_2 / X_2^r$\vspace{0.15cm}\\
&&\hspace{-0.2cm}$\sigma_{ID^*}=(sk_1,sk_2)$\vspace{0.15cm}\\
\end{tabular}
\end{minipage}
}\vspace{0.3cm}\\
\fbox{\footnotesize
\begin{minipage}[t]{1\columnwidth}
\centering {\normalsize \textbf{Query Subscription}} \vspace{0.2cm}\\
\begin{tabular}{lcl}
\hspace{-0.05cm} \underline{\small SP}\vspace{0.1cm} && \hspace{-0.2cm}\underline{\small $\Q(ID^*,\sigma_{ID^*}=(sk_1,sk_2))$}\vspace{0.2cm}\\
& &\hspace{-0.2cm}$Z_1^*=e(h, sk_1)$\vspace{0.15cm}\\
& &\hspace{-0.2cm}$Z_2^*=e(h, sk_2)$\vspace{0.15cm}\\
& &\hspace{-0.2cm}$T^*=H_2(ID^*,h,Z_1^*,Z_2^*)$\vspace{-0.1cm}\\
&\vspace{-0.05cm}\\
&\hspace{-0.3cm} $\xymatrix@1@=60pt{& \ar[l]^*+<4pt>{}_*+<4pt>{(\Q,T^*)}}$ &\vspace{-0.05cm}\\
\hspace{-0.2cm}  Store $(\Q,T^*)$ in $\mathcal{S}$\hspace{-0.2cm}& &\hspace{-0.2cm}Store $(T^*,ID^*,Z_1^*,Z_2^*)$ in $\mathcal{S}_{\Q}$\vspace{0.15cm}\\
\end{tabular}
\end{minipage}
}%
\caption{Query Registration in PEPSI. Common inputs are $q,\G,\G_T,g,h,e,\M,X_1,X_2,H_1,H_2,H_3$.}
\label{fig:ibe1usenix}
\end{figure}

\begin{table*}[t]
\centering
\subfloat[t][Actors]{
\begin{tabular}{|l|l|}
\multicolumn{1}{c}{\vspace{1.35cm}}\\
\cline{1-2}
$RA$    & Registration Authority\\
$\Q$    & Querier\\
$SP$    & Service provider\\
$\N$    & Mobile Node\\
\cline{1-2}
\end{tabular}
}
\vspace{0.5cm}\qquad
\subfloat[][Cryptographic Parameters]{
\begin{tabular}{|l|l|}
\cline{1-2}
$\lambda$   & Security parameter\\
$q,\ \G,\ \G_T,\ g,\ e$ & Public parameters\\
$\M$    & Message space\\
$H_1,\ H_2,\ H_3$           & Hash functions\\
$x_1,\ x_2$   & Registration Authority's master secret key\\
$X_1,\ X_2$ & Registration Authority's public key\\
$ID,\ ID^*$          & Query or report identifier\\
$\mathcal{D}$   & Report measurement.\\
\cline{1-2}
\end{tabular}
}
\vspace{-0.2cm}
\caption{Notation in PEPSI.\label{tab:usedsymbols1}}
\end{table*}

\subsection{Algorithms Specification}\label{sec:alg}
We now present the details of our construction.
Table~\ref{tab:usedsymbols1} defines symbols used throughout the paper.
In each protocol, we assume a secure, authenticated channel between communicating parties.

\alg{Setup.}
The Registration Authority, given a security parameter $\lambda$, generates
two groups $\G,\G_T$ of prime order $q$ with generator $g$, and a bilinear map $e: \G\times \G \rightarrow \G_T$.
Next, it picks random $x_1,x_2 \in {\F_q^*}$ and sets $X_1=g^{x_1}, X_2=g^{x_2}$.
Also, a nonce $z \in {\F_q^*}$ is selected and $h$ is set such that $h = g^z$.
The RA also defines three cryptographic hash functions $H_1: \{0,1\}^* \rightarrow \G$,
$H_2:\{0,1\}^* \rightarrow \{0,1\}^{\lambda}$ and $H_3:\{0,1\}^* \rightarrow \{0,1\}^{\lambda}$.
Secret master key is set to $msk=(x_1,x_2)$, while public key is set to $pk=(q,\G,\G_T,g,h,e,\M,X_1,X_2,H_1,H_2,H_3)$.

\alg{MN Registration.}
MN Registration is run between a mobile node \N\ and RA.
The former receives pair $(ID,z)$ where $ID$ identifies the nature of the readings for which \N\ will provides reports and $z$ is the nonce generated by RA during setup.

\alg{Query Registration.}
Query Registration is split in (i) Query Authorization, when
a querier is authorized by the RA to subscribe to a query of interest and (ii)
Query Subscription, when the querier actually subscribes to receive measurements
from the SP.
Both protocols are shown in Figure \ref{fig:ibe1usenix}.

During Query Authorization, querier \Q\ chooses an arbitrary query identifier $ID^*$, hashes it and
blinds it using some value $r$ taken uniformly at random for $\F_q^*$.
The result is sent to the RA that uses her master secret key $(x_1,x_2)$ to compute
$\mu'_1=(H_1(ID^*)\cdot g^r)^{x_1}$ and $\mu'_2=(H_1(ID^*)\cdot g^r)^{x_2}$.
Finally, \Q\ receives $(\mu'_1,\mu'_2)$ and removes the binding factor, to obtain $\sigma_{ID^*}=(sk_1,sk_2)$.

During Query Subscription, querier \Q\ uses $\sigma_{ID^*}$ to compute a tag $T^*$ that defines
her interest at the SP. Both parties then store bookkeeping information on this transaction in their respective databases.

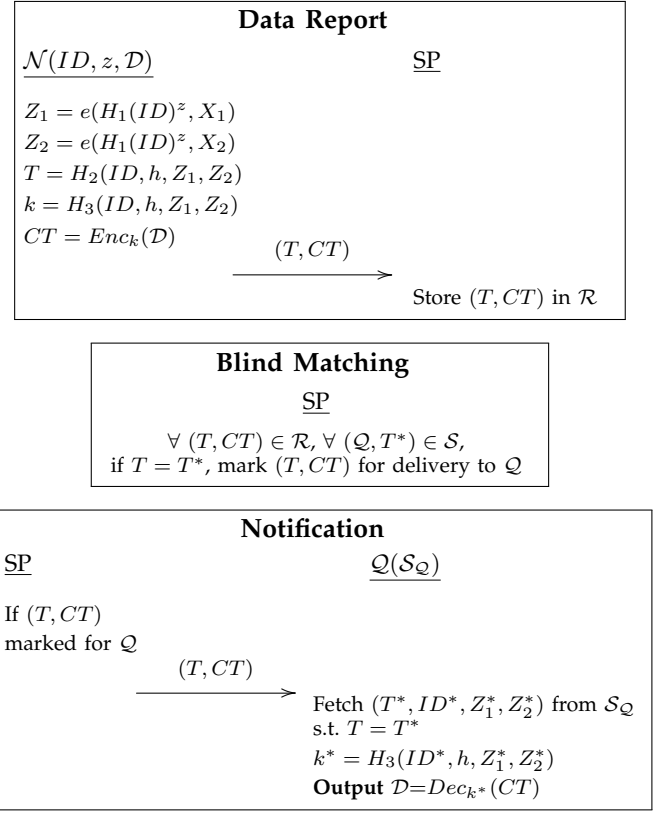
\begin{figure}[t!]
\centering
\fbox{\footnotesize
\begin{minipage}[t]{0.88\columnwidth}
\centering {\normalsize \textbf{{Data Report}}} \vspace{0.2cm}\\
\begin{tabular}{lcl}
\hspace{-0.3cm} \underline{\small $\N(ID,z,\mathcal{D})$} \vspace{0.1cm} && \hspace{-0.2cm}\underline{\small SP} \vspace{0.2cm}\\
\hspace{-0.3cm} $Z_1=e(H_1(ID)^z,X_1)$\vspace{0.1cm}\\
\hspace{-0.3cm} $Z_2=e(H_1(ID)^z,X_2)$\vspace{0.1cm}\\
\hspace{-0.3cm} $T=H_2(ID,h,Z_1,Z_2)$\vspace{0.1cm}\\
\hspace{-0.3cm} $k = H_3(ID,h,Z_1,Z_2)$\vspace{0.1cm}\\
\hspace{-0.3cm} $CT = Enc_{k}(\mathcal{D})$\vspace{-0.3cm}\\
&\hspace{-0.72cm} $\xymatrix@1@=60pt{\ar[r]^*+<4pt>{(T,CT)}&}$\\
&&\hspace{-0.2cm}Store $(T,CT)$ in $\mathcal{R}$
\end{tabular}
\end{minipage}
}%
\vspace{0.3cm}\\
\fbox{\footnotesize
\begin{minipage}[t]{0.65\columnwidth}
\centering {\normalsize \textbf{{Blind Matching}}} \vspace{0.2cm}\\
\underline{\small SP}\vspace{0.2cm}\\
$\forall\ (T,CT)\in\mathcal{R}$, $\forall\ (\Q,T^*)\in\mathcal{S}$,\\
if $T=T^*$, mark $(T,CT)$ for delivery to \Q
\end{minipage}
}\vspace{0.3cm}\\

\fbox{\footnotesize
\begin{minipage}[t]{0.96\columnwidth}
\centering {\normalsize \textbf{{Notification}}} \vspace{0.2cm}\\
\begin{tabular}{lcl}
\hspace{-0.2cm} \underline{\small SP}\vspace{0.1cm} && \hspace{0.4cm}
\underline{\small $\Q(\mathcal{S}_{\Q})$}\vspace{0.2cm}\\
\hspace{-0.2cm} If $(T,CT)$\vspace{0.07cm} \\
\hspace{-0.2cm} marked for \Q\vspace{-0.15cm}\\
&\hspace{-0.6cm} $\xymatrix@1@=60pt{\ar[r]^*+<4pt>{(T,CT)}&}$\vspace{-0.15cm}\\
&&\hspace{-0.35cm} Fetch $(T^*,ID^*,Z_1^*,Z_2^*)$ from $\mathcal{S}_{\Q}$\\
&&\hspace{-0.35cm} s.t. $T=T^*$\vspace{0.08cm}\\
&& \hspace{-0.35cm} $k^*=H_3(ID^*,h,Z_1^*,Z_2^*)$\vspace{0.08cm}\\
&& \hspace{-0.35cm} {\bf Output} $\mathcal{D}\hspace{-0.08cm}=\hspace{-0.08cm}Dec_{k^*}(CT)$
\end{tabular}
\end{minipage}
}
\caption{PEPSI online operations. Common inputs are $q,\G,\G_T,g,h,e,\M,X_1,X_2,H_1,H_2,H_3$.}
\label{fig:ibe2usenix}
\end{figure}

\alg{Data Report.}
Mobile node \N\ periodically submits data reports to SP, using the network infrastructure.
As shown in the top box of Figure \ref{fig:ibe2usenix},
reporting measurement \D\ related to query $ID$, requires the mobile node to upload a cryptographic tag $T$ and the encrypted measurement $CT$.
$T$ is computed using pair $(ID,z)$, received by the mobile node during MN Registration.
The latter is also used (through a different hash function) to derive $k$, an encryption key used to encrypt \D\ via a semantically secure symmetric encryption scheme (e.g., AES).

\alg{Query Execution.}
Query Execution is split between Blind Matching and Notification.
They are shown in the middle and lower boxes of Figure \ref{fig:ibe2usenix}, respectively.

Blind Matching involves only the SP and requires no cryptographic operations.
The SP only needs to check cryptographic tags received by mobile nodes within data reports,
against those received by queriers during their subscriptions. Any time there is a match,
the data report is marked for delivery to the subscribed querier.

Notification happens when the querier receives a new data report $(T,CT)$ that matches her interest.
At this time, querier \Q\ retrieves from her subscription database the record $(T^*,ID^*,Z_1^*,Z_2^*)$ such that $T=T^*$. Hence, \Q\ computes key $k^*$ so that $CT$ can be decrypted.

\alg{Nonce renewal.}
We assume a dynamic set of subscribed MNs where new sensors can register and malicious ones are evicted.
In order to ban misbehaving mobile nodes, the Registration Authority periodically generates and distributes a fresh $z$ to sensors
and updates the public key element $h=g^z$.
The former can be securely distributed to honest mobile nodes, e.g., using broadcast encryption \cite{broadcast}.

\section{Privacy Analysis}\label{sec:pri-an}
In this section, we formally analyze the privacy properties of PEPSI. Observe that PEPSI's
intuition lies in the application of a key-private IBE system---where query identifiers are used as identities---to protect privacy in the participatory sensing setting.
Therefore, its privacy requirements rely on the security and the key-privacy of the underlying IBE system.

As discussed above, we assume secure and authenticated channels so that the system is immune
to a wide range of attacks, e.g, session high-jacking, eavesdropping, etc.
This is a reasonable assumption as mobile nodes access participatory sensing applications using 3G networks, thus,
communication between other parties (e.g., between querier and SP) can rely on standard tools such as, TSL/SSL.

\subsection{Soundness}
PEPSI is {\em sound} if, at the end of {\em Query Execution}, querier \Q\ outputs \D, given that:

\noindent \hspace*{0.3cm}(1) \Q\ registered query $ID^*$ to the RA.\\
\hspace*{0.3cm}(2) $\exists$ node \N\ such that \N\ reports $(ID,\D)$.\\
\hspace*{0.3cm}(3) $ID^*=ID$.

Our PEPSI instantiation is {\em sound}, since, for any registered query $(ID^*,\sigma_{ID^*})$ held by querier \Q, and query identifier $ID$ reported by a node \N\, if: (1) $\sigma_{ID^*} =  (sk_1,sk_2)$,
and (2) $ID^*=ID$, we obtain:
\begin{eqnarray}
T& = & H_2[ID, h, Z_1, Z_2] = \nonumber\\
 & = & H_2[ID, g^z,e(H_1(ID)^z, X_1), e(H_1(ID)^z, X_2)]=\nonumber\\
 & = & H_2[ID^*, h, e(h,sk_1), e(h,sk_2)] = T^*\nonumber
\end{eqnarray}
and, similarly, also $k=k^*$.
Therefore, (i) SP correctly matches \Q's (authorized) subscription $T^*$
with the appropriate node report $(T, CT)$, and (ii) \Q\ can correctly decrypt $CT$ and
recover \D. $\Box$

\subsection{Node Privacy}
Informally, PEPSI is {\em node-private} if neither the NO, the SP, nor any unauthorized querier, learn any information about the type of reading or the data reported by the MN.
Also, other mobile nodes in the infrastructure should not learn any information about a given node's reports.
That is, only queriers in possession of the corresponding authorization obtain MN's readings.

Formally, privacy of node \N, providing measurement $(ID,\D)$, is guaranteed if no information about $(ID,\D)$ is leaked to unauthorized parties.

We distinguish between node privacy w.r.t. the NO and the SP and w.r.t. unauthorized queriers.

\descr{Node Privacy w.r.t. the NO and the SP.}
Privacy is considered as the probabilistic advantage that an adversary \adv\ gains
from obtaining encrypted reports. We say that PEPSI is {\em node-private} w.r.t. the NO/SP
if no polynomially bounded adversary \adv\ can win the following game with non-negligible probability above $\frac{1}{2}$.
The game is between \adv\ and a challenger $Ch$:
\begin{enumerate}
\item $Ch$ executes setup operations and computes public parameters $(q,\G,\G_T,g,h,e,\M,X_1,X_2,H_1,H_2,H_3)$ and private parameters $(msk,z)$.
\item \adv, on input public parameters, selects two reports $((ID_0,\D_0),(ID_1,\D_1))$ and gives them to $Ch$.
\item $Ch$ picks a random bit $b \in_R \{0,1\}$ and interacts with \adv\ executing the role of the node \N, on input the public parameters and private input $(ID_b,\D_b)$.
\item \adv\ outputs $b'$ and wins if $b'=b$.
\end{enumerate}

Assuming that the underlying IBE system is blind and anonymous (such as the scheme we use from~\cite{meiklejohn11usenix}),
PEPSI is trivially node-private w.r.t. the NO and the SP in the Random Oracle Model (ROM).
Assuming that $H_2$ and $H_3$ are modeled as a random oracle,
if our claim is not true then there exists a polynomial-bounded adversary $\mathcal{B}$ that breaks the CPA-security of IBE. $\Box$

\descr{Node Privacy w.r.t. unauthorized queriers.}
Privacy is considered as the probabilistic advantage that an adversary \adv\ gains
from submitting queries to the SP. We say that PEPSI is {\em node-private} w.r.t. unauthorized queriers
if no polynomially bounded adversary \adv\ can win the following game with non-negligible probability above $\frac{1}{2}$.
The game is between \adv\ and a challenger $Ch$:

\begin{enumerate}
\item $Ch$ executes setup operations and computes public parameters $(q,\G,\G_T,g,h,e,\M,X_1,X_2,H_1,H_2,H_3)$ and private parameters $(msk,z)$.
\item \adv, on input the public parameters, adaptively queries $Ch$ a number $n$ of times on a set of identifiers $L = \{ID_0, \ldots, ID_n\}$.
    For every $ID_i$, $Ch$ responds by giving \adv\ a signature $\sigma_i = (sk_{i:1}, sk_{i:2})$.
\item \adv\ announces two new identifier strings $(ID^*_0, ID^*_1) \notin L$ and generates a data record $\D^*$.
\item $Ch$ picks a random bit $b \in_R \{0,1\}$ and interacts with \adv\ taking on the role of a mobile node \N, on input the public parameters and private input $(ID^*_b,\D^*)$.
\item \adv\ outputs $b'$ and wins if $b'=b$.
\end{enumerate}

Assuming that the underlying IBE system is blind and anonymous (such as the scheme we use from~\cite{meiklejohn11usenix}),
the resulting PEPSI scheme is trivially {\em node-private} w.r.t. unauthorized queriers, in the random oracle model.
Indeed, to win the above game, \adv\ needs to forge signature on $ID^*_0$ or $ID^*_1$.
Again, if this happens, then there exists a polynomial-bounded adversary $\mathcal{B}$ that breaks the security of IBE. $\Box$

\descr{Remark.} Observe that the RA could use its secret key, $msk$, to ``test'' an arbitrary $ID^*$ against a report $(T, CT = ENC_k(\D))$.
That is, the RA could learn whether $ID^*=ID$ and violate node privacy.
However, assuming that reports $(T, CT)$ are super-encrypted under SP's public key,
the RA can access nodes' reports only if it colludes with the SP.

\subsection{Query Privacy}
Informally, PEPSI is {\em query-private} if neither the RA, the NO, the SP, other queriers, nor
any mobile node, learn any information about query interests of a querier \Q.
Query privacy with respect the RA follows from a similar argument in \cite{jarecki10scn}, that is, value $\mu$
received by the RA during Query Authorization is a random value in $\G_1$.
Query privacy w.r.t. the NO, any mobile node, and any other querier, is trivially guaranteed
as none of them obtains any cleartext message from \Q. Thus, we focus on privacy against
a malicious SP, described as the probabilistic advantage that SP gains from obtaining
queries.

Formally, PEPSI is {\em query-private} if no polynomially bounded adversary \adv\ can win the
following game with probability non-negligibly over $\frac{1}{2}$.
The game is between \adv\ and a challenger $Ch$:
\begin{enumerate}
\item $Ch$ executes setup operations and computes public parameters $(q,\G,\G_T,g,h,e,\M,X_1,X_2,H_1,H_2,H_3)$ and private parameters $(msk,z)$.
\item \adv, on input public parameters, chooses two strings $ID^*_0,ID^*_1$.
\item $Ch$ picks a random bit $b \in \{0,1\}$ and interacts with \adv\ playing the role of the querier on input the public parameters and private input $(ID^*_b)$.
\item \adv\ outputs $b'$ and wins if $b'=b$.
\end{enumerate}

Assuming that the underlying IBE system is blind and anonymous (\cite{meiklejohn11usenix})
PEPSI is trivially query-private
in the Random Oracle Model (ROM). Assuming that $H_2$ and $H_3$ are modeled as a random oracle,
if our claim is not true then there exists a polynomial-bounded adversary $\mathcal{B}$ that breaks the CPA-security of IBE. $\Box$

\subsection{Anonymity, Report Unlinkability and Location Privacy}

We do not guarantee anonymity, report unlinkability and location privacy with respect to the network operator, given the intrinsic nature of the underlying cellular network.
As discussed in Section \ref{sec:requirements}, such properties are impossible to provide w.r.t. the NO, since the NO  knows phone's position at any time.

Nonetheless, it is possible to modify our protocol to provide report unlinkability and location privacy w.r.t. all other parties, if we assume that the NO removes privacy-sensitive
metadata from each report (such as mobile nodes' identifiers, the cell from which the report was originated, etc.), before forwarding it to the SP.
Note that this would not need the use of MixNetworks, i.e., the NO does not have to delay message forwarding (e.g., until ``enough'' reports to protect privacy are collected)
but forwards ``the payload'' of each report (i.e., $(T, CT)$) as soon as it is received.

\subsection{Trust Assumptions and Limitations}

\descr{Honest-but-curious Model.}
The security of PEPSI relies on the assumption that the SP adheres to the honest-but-curious adversarial model;
that is, the server faithfully follows all protocols, but it may attempt to passively violate our privacy goals.
Observe that prior work on participatory sensing~\cite{mobisys08,comcom,dua2009,gilbert2010} assumes the presence of {\em several} non-colluding and/or fully-trusted parties.

Although it is part of future work extending our cryptographic protocols to support arbitrarily malicious adversary.
If one relaxes the honest-but-curious assumption, then it would become possible for
a malicious SP to, e.g., launch DoS attacks and disrupt the service (e.g., retain reports from queriers).
It could also collude with the RA and ``unblind'' a subscription or a report.
Finally, a malicious SP might create spurious accounts, obtain query authorizations for arbitrary identifiers, and match them against queriers/mobile nodes tags, in order to break queriers and mobile nodes privacy.

Nonetheless, we argue that assuming an honest-but-curious SP is realistic in our model, since, in participatory sensing,
SPs often capitalize on the services they provide, thus, they have no incentive to deviate from an honest-but-curious behavior.
Since the SP provides the most valuable service to large numbers of users, it has a valuable reputation to maintain and any
evidence or suspicion of misbehavior would result in a significant loss of trust and exodus of users.
Finally, we highlight that, on a similar note, many state-of-the-art privacy-enhancing technologies
in, e.g., cloud-computing environments, also assume the server provider to be honest-but-curious \cite{cao11infocom,raykova09ccsw,kamara2012salus}.

\descr{Reading trustworthiness and node reputation.}
In the current version of PEPSI, queriers do not
associate reputation or trust measures to nodes' readings, e.g., to filter
misbehaving nodes.
Although orthogonal to our main focus, i.e., privacy and confidentiality protection (which our work is the first to provably guarantee),
we readily acknowledge that we leave this as part of future work.
Nonetheless, as valid subscribers know the identity of a mobile node providing a reading of interest,
we foresee no fundamental obstacle in integrating
reputation frameworks to address reliability and trust, e.g,~\cite{wangartsense,huang2010you}.
Finally, observe that, in the semi-honest model, nodes are assumed not
to report false readings.

\descr{Repeated Readings.} PEPSI allows a querier to query all readings
satisfying certain conditions, e.g., ``temperature in San Francisco''.
While, in many scenarios, this flexibility
provides a convenient fine-grained access policy, in others, this
may require users to filter out or average, a relatively large
number of reports. However, observe that average of all measurements
can be obtained by using an (additively) homomorphic encryption scheme:
both the node reporting the measurement and the intended
recipient would derive the secret key of the homomorphic encryption
scheme. Reporting node would also compute the corresponding public
key, under which measurements are encrypted. Therefore, ciphertexts
would be homomorphic and could be aggregated at the service provider.

\section{Oprf-based Pepsi}\label{sec:oprfpepsi}

PEPSI leverages IBE to protect the privacy of both mobile nodes
and queriers during data report and query execution.
Nevertheless, pairing-based operations might be prohibitive for devices
with very limited resources or in application scenarios with high report rates. (Our performance evaluation
is presented in Section~\ref{sec:perf}.)

In this section, we introduce another instantiation of PEPSI that avoids bilinear map operations.
The main idea is to design a new Query Authorization protocol, replacing the blind IB signature scheme with OPRF.
PEPSI can use \emph{any} OPRF protocol, however,
in this paper, we use a blind-RSA signature based OPRF that is secure
under the one-more RSA assumption \cite{bellare2003one}.
In particular, the OPRF is defined as $f_d(x)=H'(H(x)^d)$ where $H(\cdot)$ and
$H'(\cdot)$ are defined as random oracles and $d$ is the secret RSA exponent of the signer.

During Query Authorization, the RA acts as the sender with private input its RSA signing key,
while querier \Q\ acts as the receiver with private input an arbitrary query identifier $ID^*$.
At the end of the protocol, \Q\ obtains the RA signature on $ID^*$, while
RA learns nothing.
Hence, the signature is used to compute the cryptographic tag
$T^*$ uploaded to the SP during Query Subscription.
Similarly, mobile nodes must receive a signature by the Registration Authority on arbitrary query identifier $ID$,
before reporting measurements.
This might be achieved as in the MN Registration protocol of Section \ref{sec:alg},
or it could leverage a ``blind'' protocol as for Query Authorization; in the latter case,
mobile nodes would keep the nature of their reports private w.r.t. the RA.
In any case, the signature obtained by the mobile node is used to compute the cryptographic tag $T$ that is sent along with each report and the encryption
key used to encrypt the measurement.

Details of OPRF-based PEPSI protocols are provided below. We partially re-use the notation of Table~\ref{tab:usedsymbols1}, however,
we denote with $d$, resp.,$(N,e)$, RA's master secret key, resp., public key.

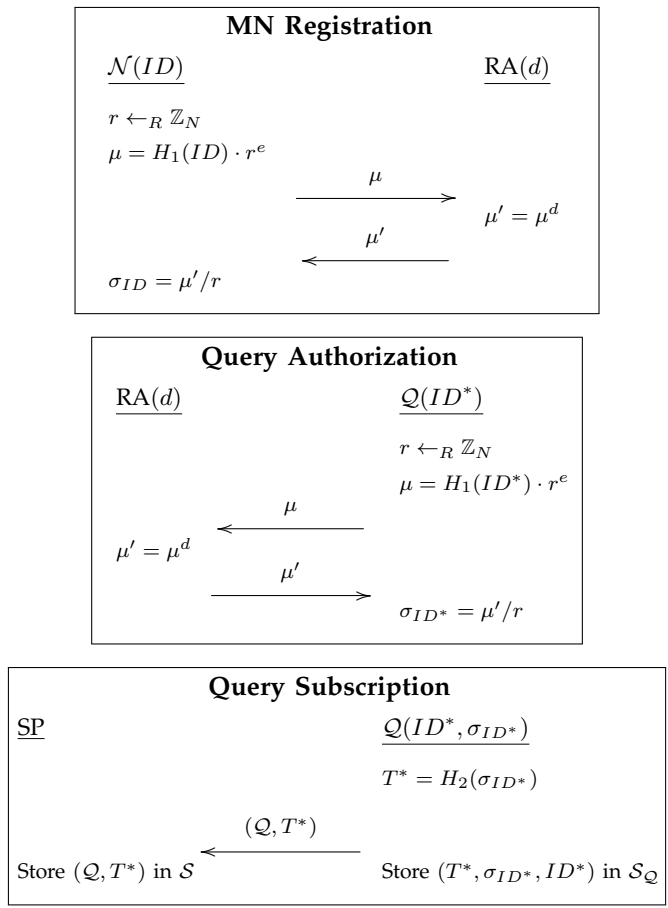
\begin{figure}[t!]
\centering
\fbox{\footnotesize
\begin{minipage}[t]{0.75\columnwidth}
\centering {\normalsize \textbf{MN Registration}} \vspace{0.2cm}\\
\begin{tabular}{lcl}
\hspace{-0.2cm} \underline{\small \N$(ID)$}\vspace{0.1cm} && \hspace{-0.2cm}\underline{\small RA$(d)$}\vspace{0.2cm}\\
\hspace{-0.2cm} $r \leftarrow_R \Z_{N}$\vspace{0.15cm}\\
\hspace{-0.2cm} $\mu = H_1(ID)\cdot r^{e}$\vspace{-0.1cm}\\
&\hspace{-0.2cm} $\xymatrix@1@=60pt{\ar[r]^*+<4pt>{\mu}&}$ \vspace{-0.1cm}\\
&&  \hspace{-0.2cm}$\mu'=\mu^{d}$ \hspace{-0.4cm}\vspace{-0.2cm}\\
&\hspace{-0.3cm} $\xymatrix@1@=55pt{& \ar[l]^*+<4pt>{}_*+<4pt>{\mu'}}$ &\vspace{-0.05cm}\\
\hspace{-0.2cm} $\sigma_{ID}=\mu'/r$\vspace{0.15cm}\\
\end{tabular}
\end{minipage}
}\vspace{0.3cm}\\
\fbox{\footnotesize
\begin{minipage}[t]{0.70\columnwidth}
\centering {\normalsize \textbf{Query Authorization}} \vspace{0.2cm}\\
\begin{tabular}{lcl}
 \underline{\small RA$(d)$}\vspace{0.1cm} && \hspace{-0.2cm}\underline{\small $\Q(ID^*)$}\vspace{0.2cm}\\
&&\hspace{-0.2cm}$r \leftarrow_R \Z_{N}$\vspace{0.15cm}\\
&&\hspace{-0.2cm}$\mu = H_1(ID^*)\cdot r^{e}$\vspace{-0.1cm}\\
&\hspace{-0.3cm} $\xymatrix@1@=55pt{& \ar[l]^*+<4pt>{}_*+<4pt>{\mu}}$ &\vspace{-0.05cm}\\
 $\mu'=\mu^{d}$ \hspace{-0.4cm}\spa && \vspace{-0.2cm}\\
&\hspace{-0.2cm} $\xymatrix@1@=60pt{\ar[r]^*+<4pt>{\mu'}&}$ \vspace{-0.1cm}\\
&&\hspace{-0.2cm}$\sigma_{ID^*}=\mu'/r$\vspace{0.15cm}\\
\end{tabular}
\end{minipage}
}\vspace{0.3cm}\\
\fbox{\footnotesize
\begin{minipage}[t]{0.95\columnwidth}
\centering {\normalsize \textbf{Query Subscription}} \vspace{0.2cm}\\
\begin{tabular}{lcl}
\hspace{-0.2cm}\underline{\small SP}\vspace{0.1cm} && \hspace{-0.2cm}\underline{\small $\Q(ID^*,\sigma_{ID^*})$}\vspace{0.2cm}\\
& &\hspace{-0.2cm}$T^*=H_2(\sigma_{ID^*})$\vspace{-0.1cm}\\
&\vspace{-0.05cm}\\
&\hspace{-0.5cm} $\xymatrix@1@=60pt{& \ar[l]^*+<4pt>{}_*+<4pt>{(\Q,T^*)}}$ &\vspace{-0.05cm}\\
\hspace{-0.2cm}Store $(\Q,T^*)$ in $\mathcal{S}$\hspace{-0.2cm}& &\hspace{-0.2cm}Store $(T^*,\sigma_{ID^*},ID^*)$ in $\mathcal{S}_{\Q}$\vspace{0.15cm}\\
\end{tabular}
\end{minipage}
}%
\caption{OPRF-based PEPSI -- offline operations. Common inputs are $N,e,H_1,H_2,H_3$.}
\label{fig:oprf1}
\end{figure}

\subsection{Protocols Description}

\alg{Setup.}
The Registration Authority (RA), on input of the security parameter $\lambda$, generates a so-called safe RSA modulus $N=pq$, where $p$ and $q$ are random distinct
$\lambda_1$-bit primes, (where $\lambda_1$ is a polynomial function of $\lambda$),
such that $p=2p'+1$ and $q=2q'+1$ for distinct primes $p',q'$.
Fixed $N$, the RA picks a random positive integer $e<\phi(N)$ such that
$\gcd(e,\phi(N))=1$, and later computes $d$ such that $ed=1 \bmod{\phi(N)}$.
It also picks a Full Domain Hash (FDH) function $H_1: \{0,1\}^* \rightarrow \Z_N$
and two additional hash functions $H_2: \{0,1\}^{\Z_N} \rightarrow \{0,1\}^{\lambda_2}$,
$H_3: \{0,1\}^{\Z_N} \rightarrow \{0,1\}^{\lambda_3}$ (where both $\lambda_2,\lambda_3$ are polynomial functions of $\lambda$).
The RA keeps $d$ private and publishes $N,e,H_1,H_2,H_3$.

\alg{MN Registration.}
As described before, \N\ might just receive $(ID,\sigma_{ID})$ from the RA,
where $ID$ identifies the type of reports and $\sigma_{ID}=H_1(ID)^d$.
Alternatively, we could preserve the privacy of reports w.r.t. the RA, defining
a MN Registration protocol based on OPRF.
The protocol is depicted in the top box of Figure \ref{fig:oprf1} and
is essentially the OPRF protocol based on blind-RSA signatures
where the mobile nodes plays the role of the receiver.
We assume that before the protocol starts, RA authenticates \N.
Hence, the latter picks a random value $r\in\Z_{N}$ and blinds its query identifier $ID$;
the blinded hash is sent to RA for signing.
The received signature is unblinded by \N\ to retrieve $\sigma_{ID}$, a valid signature of
$ID$ under the secret key of the RA.

\alg{Query Registration.}
This protocol is divided in Query Authorization and Query Subscription,
both depicted in Figure \ref{fig:oprf1}.

Query Authorization is once again  the OPRF protocol but this time the querier is playing as the receiver
on her private input $ID^*$.
At the end of the protocol, she obtains $\sigma_{ID^*}$, a valid signature of $ID^*$ under the secret key of the RA.

During Query subscription, querier \Q\ computes a cryptographic tag $T^*$ from $\sigma_{ID^*}$
and uploads it at SP. From that moment on, SP forwards to \Q\ all reports that match her
interest.

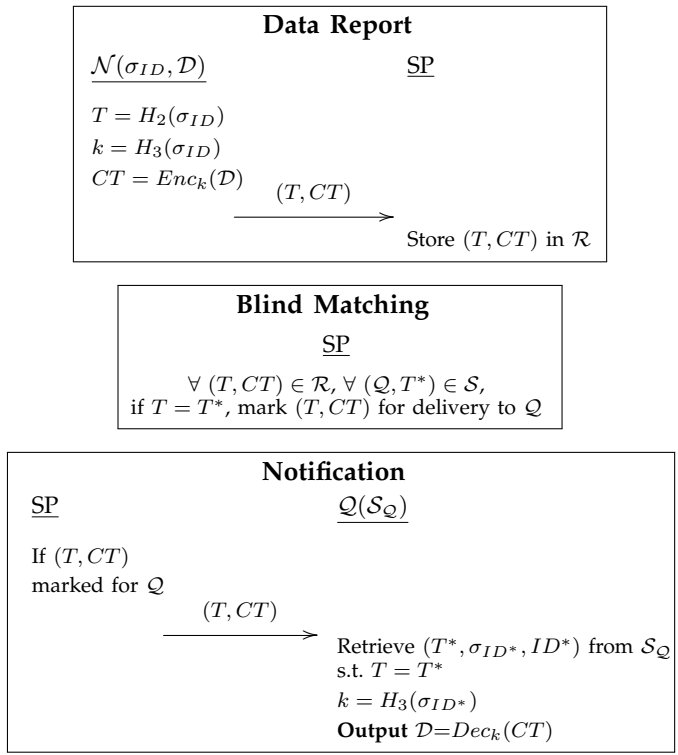
\begin{figure}[t!]
\centering
\fbox{
\footnotesize
\begin{minipage}[t]{0.75\columnwidth}
\centering {\normalsize \textbf{{Data Report}}} \vspace{0.2cm}\\
\begin{tabular}{lcl}
\hspace{-0.3cm} \underline{\small $\N(\sigma_{ID},\D)$} \vspace{0.1cm} &&  \hspace{-0.3cm}\underline{\small SP} \vspace{0.2cm}\\
\hspace{-0.3cm} $T = H_2(\sigma_{ID})$\vspace{0.1cm}\\
\hspace{-0.3cm} $k = H_3(\sigma_{ID})$\vspace{0.1cm}\\
\hspace{-0.3cm} $CT = Enc_{k}(\D)$\vspace{-0.3cm}\\
&\hspace{-0.72cm} $\xymatrix@1@=60pt{\ar[r]^*+<4pt>{(T,CT)}&}$\\
&&\hspace{-0.3cm}Store $(T,CT)$ in $\mathcal{R}$
\end{tabular}
\end{minipage}
}%
\vspace{0.3cm}\\
\fbox{\footnotesize
\begin{minipage}[t]{0.63\columnwidth}
\centering {\normalsize \textbf{{Blind Matching}}} \vspace{0.2cm}\\
\underline{\small SP}\vspace{0.2cm}\\
$\forall\ (T,CT)\in\mathcal{R}$, $\forall\ (\Q,T^*)\in\mathcal{S}$,\\
if $T=T^*$, mark $(T,CT)$ for delivery to \Q
\end{minipage}
}\vspace{0.3cm}\\
\fbox{\footnotesize
\begin{minipage}[t]{0.965\columnwidth}
\centering {\normalsize \textbf{{Notification}}} \vspace{0.2cm}\\
\begin{tabular}{lcl}
\underline{\small SP}\vspace{0.1cm} &&
\hspace{-0.32cm} \underline{\small $\Q(\mathcal{S}_{\Q})$}\vspace{0.2cm}\\
 If $(T,CT)$\vspace{0.07cm} \\
 marked for \Q\vspace{-0.15cm}\\
&\hspace{-0.6cm} $\xymatrix@1@=58pt{\ar[r]^*+<4pt>{(T,CT)}&}$\vspace{-0.15cm}\\
&&\hspace{-0.32cm} Retrieve $(T^*,\sigma_{ID^*},ID^*)$ from $\mathcal{S}_{\Q}$\\
&&\hspace{-0.32cm} s.t. $T=T^*$\vspace{0.08cm}\\
&& \hspace{-0.32cm} $k=H_3(\sigma_{ID^*})$\vspace{0.08cm}\\
&& \hspace{-0.32cm} {\bf Output} $\D\hspace{-0.08cm}=\hspace{-0.08cm}Dec_k(CT)$
\end{tabular}
\end{minipage}
}
\caption{OPRF-based PEPSI - online operations. Common inputs are $N,e,H_1,H_2,H_3$.}
\label{fig:oprf2}
\end{figure}

\alg{Data Report.}
Protocols details are provided in the upper box of Figure \ref{fig:oprf2}.
When mobile node \N\ wants to report measurement $\mathcal{D}$ related to
query $ID$, it computes a cryptographic tag using R$\sigma_{ID}$.
The latter is also used to compute a key of a symmetric encryption scheme.
Finally, the encrypted data report and the cryptographic tag are uploaded at SP.

\alg{Query Execution}
This protocol is divided in Blind Matching and Notification, that are
depicted in the middle and bottom boxes of Figure \ref{fig:oprf2}, respectively.

Blind Matching only involves the SP and requires no cryptographic operations.
The SP only needs to find all received reports that match a subscription.
A report $(T,CT)$ matches a subscription $(\Q, T^*)$ if $T=T^*$.
In this case, the report is marked for delivery to \Q.

During Notification, SP sends to \Q\ all reports that match her subscription(s).
For each received report $(T,CT)$, \Q\ retrieves the tuple $(T^*,\sigma_{ID^*},ID^*)$ such
that $T=T^*$. Hence, \Q\ uses $\sigma_{ID^*}$ to derive key $k^*$ and uses the latter
to decrypt ciphertext $CT$.

\subsection{Discussion}

The OPRF-based version of PEPSI is trivially sound.
Its security (i.e., node privacy and query privacy) relies on the security of the underlaying OPRF (namely, blind-RSA signatures).
That is, an adversary that is capable of breaking the privacy of either a mobile node or a querier can be used to
build another adversary capable of breaking the security of the OPRF protocol.

While we use blind-RSA signatures as building block, PEPSI could actually leverage any OPRF.
If new (e.g., faster) OPRF constructions are available, they could be seamlessly integrated in PEPSI for better performance.

The IBE-based version of PEPSI separates roles of mobile nodes and queries.
On one side, once a mobile node registers for arbitrary query $ID$, it is allowed to provide reports
but cannot subscribe for other mobile node reports on the same $ID$.
On the other side, queriers that subscribe for arbitrary query $ID^*$, cannot provide reports for the same query.
In the OPRF-based version of PEPSI, the set of queries and mobile nodes for a particular query is not necessarily disjoint, i.e.,
once obtained a signature by the RA for query $ID$, it is possible either to provide reports or to subscribe for measurements related to $ID$.
Nevertheless, participatory sensing communities are constituted by users that interchangeably provide and retrieve data for the
same phenomena (e.g., bikers that retrieve pollution data for their next routes and that provide those measurements while they are biking).
Hence, we argue that grouping of mobile nodes and queriers is a reasonable trade-off to achieve a considerable performance improvement.

\section{Performance Evaluation}\label{sec:perf}
This section provides an analytical and empirical evaluation of the cost of cryptographic operations
involved in PEPSI.
\modified{
We focus our evaluation on the overhead incurred at Mobile Nodes, since their are usually devices with constrained resources.
We do not study the performance/scalability of the Service Provider (SP). To this end, we note that a SP in PEPSI 
only needs to match tags (e.g., 160-bit binary strings) reported by Mobile Nodes against the ones uploaded by Queriers.
The Service Provider is not involved in any cryptographic operations and we argue that its performance are comparable to the ones 
of a SP in a non-private system, where publishers and subscribers upload measurements and queries in the clear.
}

\begin{table*}[t]
\footnotesize
\centering
\begin{tabular}{lcc|cc|}
\cline{2-5}
&\multicolumn{2}{|c|}{\textbf{PEPSI}} & \multicolumn{2}{|c|}{\textbf{OPRF-PEPSI}}\\
\cline{2-5}
\cline{2-5}
& \multicolumn{1}{|c}{Exp. / Mult.} &\multicolumn{1}{c|}{Map / Hash} & \multicolumn{1}{c}{Exp. / Mult.} &\multicolumn{1}{l|}{Map / Hash}\\
\hline
\multicolumn{1}{|l|}{Node registr. (\N)}   &  -  /  -  &  -  /  -       &  $1$ / $2$ & - / $1$\\
\hline
\multicolumn{1}{|l|}{Data report   (\N)}   & $1$ /  -  & $2$ / $1$      &   -  /  -  & - / $2$\\
\hline
\multicolumn{1}{|l|}{Query auth.   (\Q)}   & $1$ / $3$ &  -  / $1$      &  $1$ / $2$ & - / $1$\\
\hline
\multicolumn{1}{|l|}{Query subscr.  (\Q)}   &  -  /  -  & $2$ / $1$      &   -  /  -  & - / $1$\\
\hline
\multicolumn{1}{|l|}{Notification  (\Q)}   &  -  /  -  &  -  / $1$      &   -  /  -  & - / $1$\\
\hline
\end{tabular}
\caption{Computation overhead for mobile nodes and queriers in PEPSI and OPRF-PEPSI. Symmetric encryption and decryption are not taken into account.\label{tab:comp}}
\end{table*}

\descr{Theoretical Analysis.}
Table~\ref{tab:comp} shows the theoretical computation overhead incurred at mobile nodes and queriers for both PEPSI and OPRF-PEPSI.
In PEPSI, the most involved protocol for a mobile node is data report where the node performs one modular exp, computes one hash function
and two bilinear map evaluations. Queriers face the most complex operations during query subscription when they compute one hash functions and two
bilinear map evaluations.
OPRF-PEPSI allows for lower overhead as operations at mobile nodes or queriers only involve modular multiplications and short exponentiations, or hash function evaluations.

\descr{Experimental evaluation.}
We implemented protocol operations executed by MNs
on a Nokia N900 (equipped with a 600 MHz ARM processor and 256 MB RAM)
running {\tt libpbc}~\cite{pbc} and {\tt gmp}~\cite{gmp} cryptographic libraries. %

For the instantiation based on pairings, we selected Type-A pairings and $160$-bit prime $q$.
Computation overhead is due to the computation of $T$, the encryption key $k$, and the encrypted
report $CT$. Note that the first two values can be computed off-line, independently of the sensed data.
Communication overhead is merely due to the transmission of $T$, which is the output of a hash function (e.g., SHA-1), and can be as small as $160$-bit.
Indeed, using available symmetric-key cryptosystems (e.g., AES), the length of $CT$ is almost the same as a
reading $\mathcal{D}$.

Without leveraging off-line pre-computation, we measured the time to compute and transmit $(T,CT)$,
using integers as data reports.
Over $100$ experiments, we experienced an average time of $93.47 ms$ to compute $(T,CT)$ and
around $80 ms$ for transmission over the 3G network. Note that a \naive\ (non-private) solution
would save in computation (since data would not be encrypted) but would spend roughly the same transmission time to send the report.
Finally, remark that the SP incurs no communication nor computational overhead: its task is limited to forwarding and hash comparisons.
Similarly, the only additional operation that queriers perform during query execution is the symmetric decryption of received readings,
which incurs a negligible overhead.

OPRF-based PEPSI enjoys the same features of the protocols presented in Section \ref{sec:pepsi}
without resorting to expensive pairing operations.
In particular, the most expensive operation is the RSA signature during mobile node / querier registration.
According to our experiments, an RSA signature with a 1024-bit takes $8.4ms$ (average over $100$ experiments) using Chinese Remainder Theorem.\footnote{See items 14.71 and
14.75 in~\cite{handbook} for more details on CRT-based exponentiation.}
Other operations include modular multiplications ($0.03ms$), hashing and symmetric encryption/decryption
that have negligible overhead.

\section{Conclusion}\label{sec:conclusion}
The participatory sensing paradigm bears a great potential. However, its success depends
on the number of users willing to report measurements from their mobile devices.
Clearly, a wide-scale user participation is bound to effective protocols that preserve
privacy of both data producers (i.e., mobile nodes)  and data consumers (i.e., queriers).
In this paper, we have highlighted shortcomings of previous solutions %
and we embarked toward a formal treatment of privacy in participatory sensing.
To this aim, we analyzed which are the privacy features that can be guaranteed with provable security and
introduced two private participatory sensing instantiations that attain them.
Finally, we provided figures of the incurred overhead at mobile nodes.

PEPSI allows for privacy-preserving information dissemination in participatory sensing application.
While assessing the trustworthiness of submitted reports is not  our main focus, we stress that PEPSI can be
plugged in participatory sensing frameworks that employ trusted hardware \cite{dua2009,gilbert2010,saroiu10hotmobile}.
Alternatively, since reports and subscription are not anonymous PEPSI could be used in conjunction with
reputation systems \cite{josang07dss} to apprise user reputation and evict malicious users.

As often happens, deploying actual solutions based on our proposal requires addressing  additional (potential) security issues. \modified{As part of future work, we plan to study and analyze, in realistic use cases, trade-offs between efficiency and utility related to different levels of keywords' granularity, which we have left out of the scope of this paper.} 
Also, our efforts are focusing on reputation management, data integrity, DoS prevention, as well as Sybil attacks.
Our immediate next step is to deploy testing applications using the PEPSI infrastructure,
as well as to devise a large-scale evaluation of its global overhead.
Interesting open challenges, calling for further research in the area, remain in how to provide location privacy with respect to cellular
network operator, addressing potential collusion between different parties, integrating
trust and reputation frameworks, and
supporting more complex queries (e.g., aggregate and conjunctive queries).

\bibliographystyle{abbrv}
\bibliography{bibfile}
\balance

\end{document}